\documentclass[a4paper, showpacs, showkeys,11pt]{revtex4}
\usepackage[utf8x]{inputenc}
\usepackage{amssymb}
\usepackage{graphicx}
\usepackage{default}
\usepackage{amssymb,amsmath}
\usepackage{subfig,graphicx,caption}
\usepackage{hyperref}  
\usepackage{setspace}
\usepackage{wrapfig,sidecap,,fullpage,morefloats}
\usepackage{amsmath,amsthm,amssymb,amstext,amsopn,amsxtra}
\usepackage{latexsym}
\usepackage{default}
\newcommand{\be}{\begin{equation}}
\newcommand{\bea}{\begin{eqnarray}}
\newcommand{\bc}{\begin{center}}            
\newcommand{\ee}{\end{equation}}
\newcommand{\eea}{\end{eqnarray}}
\newcommand{\ec}{\end{center}}
\newcommand{\baa}{\begin{eqnarray*}}
\newcommand{\eaa}{\end{eqnarray*}}
\begin{document}
\title{Optimal engine performance using inference for non-identical finite source and sink}
\author{{Preety Aneja \footnote{preetyaneja@iisermohali.ac.in}}, Harsh Katyayan and 
{Ramandeep S. Johal \footnote{rsjohal@iisermohali.ac.in}}}
\affiliation{Department of Physical Sciences \\ Indian Institute of Science
 Education and Research, Mohali
 \\ Sector 81, Knowledge city \\ Manauli P.O., Ajit Garh - 140306, Punjab, India}
\begin{abstract}
We quantify the prior information to infer the optimal characteristics for a constrained thermodynamic process of maximum 
work extraction for a pair of non-identical finite systems. The total entropy of the whole system remains conserved. The 
ignorance is assumed about the final temperature of the finite systems and then a prior distribution is assigned to the 
unknown temperatures. We derive the estimates of efficiency for this reversible model of heat engine with incomplete information.
The estimates show good agreement with efficiency at optimal work for arbitrary sizes of systems, however the estimates
become exact when one of the reservoir becomes very large in comparison to the other.
\end{abstract}
\pacs{05.70.−a, 05.70.Ln, 02.50.Cw}
\keywords{Inference; Constrained thermodynamic process; Finite source/sink}
 \maketitle
\section{Introduction}
In this paper, we revisited the problem of maximum work extraction from the inference approach 
\cite{preetycejp,preetyiop,johalfphys,johaljnet,preetyarxv}. The extraction of maximum work using a pair of finite source and 
sink has been 
discussed
earlier in literature \cite{ondrechen,callen,leff,lavenda,landsberg}. The whole set-up working as a heat engine delivers maximum 
work for a reversible
process in accordance with {\it Maximum Work Theorem}. The maximum achievable efficiency of such a heat engine
is clearly is less than that of the {\it Carnot efficiency} ($\eta_c$) because of the finite size of source/sink. In 
recent years, the field of finite-time thermodynamics has become a popular area of research since it deals with 
the realistic constraints in heat engines such as finite-time of the engine operation cycle, finite reservoirs,
internal friction etc. These constraints, in turn, lead to a less efficient heat engine than Carnot engine 
but is of practical importance. However, such engines can be optimised to deliver maximum power with a 
compromise of efficiency. 

The inference approach is based on Bayesian approach to uncertainty \cite{bayes,laplace,jeffreys,jaynes} in which any uncertainty 
can be treated probabilistically. This probabilistic approach is associated with a rational degree of belief of the observer
rather than relative frequency interpretation of probability \cite{fran,welsh}. The uncertain parameter 
is assigned with a probability distribution simply known as a prior to quantify the ignorance of the likely values of the uncertain 
parameter. Priors with different justification have been proposed \cite{laplace,raiffa,jeffreys1,bernardo,Abe2014}.
In our work, we propose prior for a constrained thermodynamic process with incomplete information of the thermodynamic coordinates.

In our previous work \cite{preetyiop}, we addressed the problem of maximum work extraction with finite source/sink within this 
inference approach. A pair of identical, similar sized finite reservoirs were considered which serve as finite 
source and sink. The fundamental thermodynamic relation obeyed by the reservoirs is taken to be $S = \kappa U^{\omega_1}$,
where $\kappa$ may depend on some universal constants and/or volume, particle number of the system. We restrict to the case 
$0<\omega_1<1$, which implies systems with a positive heat capacity. The optimal or maximum work extracted from this reservoir 
set-up is estimated. The efficiency at optimal work is also inferred upto second order as $\eta \approx {\eta_c}/2 + {\eta_c}^2/8$ 
\cite{curzon,schmiedl,zc,esposito}, in near equilibrium regime ($\eta_c \ll 1$).
A generalisation of this approach can be thought of by considering the non-identical systems as reservoirs \cite{johaljnet}.
In paper \cite{johaljnet}, finite reservoirs are modelled by perfect gas systems with different constant heat capacities.
Thus, the new information about distinct source and sink was utilized in the assignment of prior for the uncertain 
temperatures. The temperatures $T_1$ and $T_2$ can be distinguished now. Moreover, the ranges of allowed values of $T_1$ and 
$T_2$ are different. In this paper, we consider two dissimilar systems obeying a thermodynamic relation of the form 
$S = \kappa U^{\omega_1}$ and reconsider the maximum work extraction process within inference approach. We derive the
temperature and efficiency estimates for this model which show remarkable agreement with their optimal
values.

This paper is organised as follows. In section \ref{model3}, we discuss the model for finite reservoirs. 
Section \ref{prior4} outlines the discussion of the range for $T_1$ and $T_2$ and the form of priors. It also
comprises of the discussion of inference for special cases when one system becomes very large in comparison to the other.
Then, we discuss the estimation procedure close to equilibrium analytically. In section \ref{nresults}, numerical
results for arbitrary sizes of reservoirs have discussed. Finally in section \ref{conc4}, we make some concluding
remarks on our extended inference approach applied in case of non-identical systems.
\section{Model}{\label{model3}}
To model the finite reservoirs, consider a pair of thermodynamic systems obeying the relation of the form $S = \kappa U^{\omega_1}$, 
where $U$ is the internal energy of the system and $\omega_1$ is some known constant. Some well-known
physical examples in this framework are the ideal Fermi gas ($\omega_1=1/2$),
the degenerate Bose gas ($\omega_1=3/5$) and the black body radiation ($\omega_1=3/4$) \cite{zylka}. 
Classical ideal gas can also be treated as the limit, $\omega_1\to 0$.
Using the basic definition : ${\partial S}/{\partial U} = {1}/{T}$, we get: $U = {(\omega_1 \kappa T)}^{1+\omega}$. 
Alternately, we can write : $S = \kappa^{1+\omega} (\omega_1 T)^{\omega}$, 
where  $\omega = \omega_1/(1-\omega_1)$.

Since the 
thermodynamic relation obeyed by the two systems remains the same, the two may be non-identical only if they differ in their
volumes, number/nature of particles etc. Thus, it is the constant of proportionality, $\kappa$, which is different for the 
two systems. Let $T_+$ and $T_-$ ($<T_+$) be the initial temperatures of two systems with $\kappa_1$ and $\kappa_2$ as the 
proportionality constants respectively.

To perform inference, examine an arbitrary intermediate stage of the process when the temperatures of the two systems are 
${T}_{1}^{'}$ 
and ${T}_{2}^{'}$ respectively. The work extracted from the engine is $W = -\bigtriangleup U$ which can be written as:
\be
W = (k_2 \omega_1)^{1+\omega}\left[\gamma^{-(1 + \omega)}({T_+}^{1+\omega}-{{T}_{1}^{'}}^{1+\omega})+({T_-}
^{1+\omega}-{{T}_{2}^{'}}^{1+\omega})\right],
\ee
where $\gamma = {\kappa_2}/{\kappa_1}$. For convenience, we define $\gamma^{1 + \omega} = \sigma$, $\theta = T_-/T_+$, 
$T_1 = {T}_{1}^{'}/T_+$ and $T_2 = {T}_{2}^{'}/T_-$.
Thus, work can be rewritten as :
\be
W = (k_2 \omega_1 T_+)^{1+\omega}\left[\sigma^{-1}(1-{{T}_{1}}^{1+\omega})+({\theta}^{1+\omega}-{{T}_{2}}
^{1+\omega})\right].
\label{work}
\ee
The constraint of entropy conservation $\bigtriangleup S = 0$ gives $S_1 + S_2 = S_+ + S_-$, where $S_+$ 
($S_-$) and $S_1$ $(S_2)$ are the values of entropy in initial and final state of system 1 (2) respectively. 
 Then, we can write:
\be
T_1 = \left[1 + \sigma({\theta}^{\omega}-{T_2}^{\omega})\right]^{\frac{1}{\omega}},
\label{T1rel}
\ee
or equivalently,
\be
T_2 = \left[{\theta}^{\omega} + \sigma^{-1}(1-{T_1}^{\omega})\right]^{\frac{1}{\omega}}.
\label{T2rel}
\ee
By making use of above equations, work can be written as a function of one variable (say $T_2$) only:
\bea
W (T_2)& = & (k_2 \omega_1 T_+)^{1 + \omega}\left[\sigma^{-1}\left(1-(1 + \sigma
({\theta}^{\omega}-{T_2}^{\omega}))^{\frac{1+\omega}{\omega}}\right)
+ \left({\theta}^{1+\omega}-{T_2}^{1+\omega}\right)\right].
\label{workT2}
\eea
Similarly, work can be written as a function of $T_1$ also.

The optimal work can be extracted from the engine when the two systems reach a common temperature. 
The common temperature, $T_c$ , is given from Eq. (\ref{T1rel}) with $T_1 = T_2 = T_c$ as:
\be
T_c = {\left(\frac{1 +\sigma {\theta}^{\omega}}{1+\sigma}\right)}^{\frac{1}{\omega}}.
\label{tcgamma}
\ee

The efficiency of the engine is given as $\eta = 1 - Q_{\rm{out}}/Q_{\rm{in}}$, where $Q_{\rm{in}} = (\omega_1 \kappa_1 T_+)
^{1+\omega} 
(1 - {T_1}^{1+\omega})$ is the heat absorbed from the source and $Q_{\rm{out}} = (\omega_1\kappa_2T_+)^{1+\omega}(T_2^{1+\omega}
-{\theta}^{1+\omega})$ is the waste heat rejected to the sink. Thus, for any arbitrary value of $\gamma$, efficiency at any 
intermediate stage of the process can be given as:
\be
\eta_{\gamma} = 1 - \sigma \frac{({T_2}^{1 + \omega}-\theta^{1 + \omega})}{(1-{T_1}^{1 + \omega})}.
\label{effg}
\ee
For efficiency at optimal work ($\eta_{\gamma}^{*}$), we substitute $T_1 = T_2 = T_c$ in above equation to obtain:
\be
\eta_{\gamma}^{*}  = 1 - \sigma \frac{({T_c}^{1 +\omega}-\theta^{1+\omega})}{(1-{T_c}^{1 + \omega})}.
\label{effop}
\ee
Let us discuss the efficiency in the limiting cases:\\
(a) In the limit $\gamma \rightarrow 0$ i.e. when heat source is very large as compared to heat sink, temperature of source
remains constant at ${T}_{1}^{'} = T_+$ or $T_1 = 1$ while temperature of sink approaches this value for optimal work extraction.
We write efficiency as a function of $T_2$ as
\be
\eta_{0} = 1 - \left(\frac{\omega}{1+\omega}\right)\left(\frac{{T_2}^{1+\omega}-{\theta}^{1 + \omega}}{{T_2}^{\omega}
-{\theta}^{\omega}}\right).
\label{effg0}
\ee
Efficiency at optimal work in this limit is given by substituting $T_2 = 1$ in above equation as:
\be
\eta_{0}^{*} =1 - \left(\frac{\omega}{1+\omega}\right)\left(\frac{1-{\theta}^{1 + \omega}}
{1-{\theta}^{\omega}}\right).
\label{effopg0}
\ee
(b) In the limit $\gamma \rightarrow \infty$ i.e. when heat sink is very large in comparison to heat source, the sink stays 
at temperature ($T_-$) and source approaches this value for optimal work extraction. The efficiency can be expressed in terms
of $T_1$ as:
\be
\eta_{\infty} = 1 - \theta \left(\frac{1+\omega}{\omega}\right)\left(\frac{1-{T_1}^{ \omega}}{1-{T_1}^{1+\omega}}\right).
\label{effgl}
\ee
For the optimal process, substitute $T_1 = \theta$ in above expression to obtain:
\be
\eta_{\infty}^{*} =1 - \theta \left(\frac{1+\omega}{\omega}\right)\left(\frac{1-{\theta}^{\omega}}
{1-{\theta}^{1+\omega}}\right).
\label{effopgl}
\ee
\section{Assignment of prior}{\label{prior4}}
The inference is performed by assigning the prior probability distributions for the uncertain parameters $T_1$ and $T_2$, 
since we assume ignorance of the actual values of $T_1$ and $T_2$ or the extent to which the process has proceeded. Thus, 
there are two observers, 1 and 2, for $T_1$ and $T_2$ respectively. Let us first summarise the prior information we possess 
before making the inference:\\
(i) There exists a one-to-one relation between $T_1$ and $T_2$ given by Eq. (\ref{T1rel}) or Eq. (\ref{T2rel}) which suggests 
that probability of $T_1$ to lie in small range [$T_1, T_1 + dT_1$] is same as the probability of $T_2$ to lie in 
[$T_2,T_2 + dT_2$], so we can write :
\be
P_1(T_1) dT_1 = P_2 (T_2) dT_2,
\label{priorrel}
\ee
where $P_1$ and $P_2$ are the normalised prior distribution functions for $T_1$ and $T_2$.\\
(ii) The set-up works like a heat engine and thus $W\geq 0$.
 
\begin{figure}[h]
\begin{tabular}{ll}
\includegraphics[scale = 0.5]{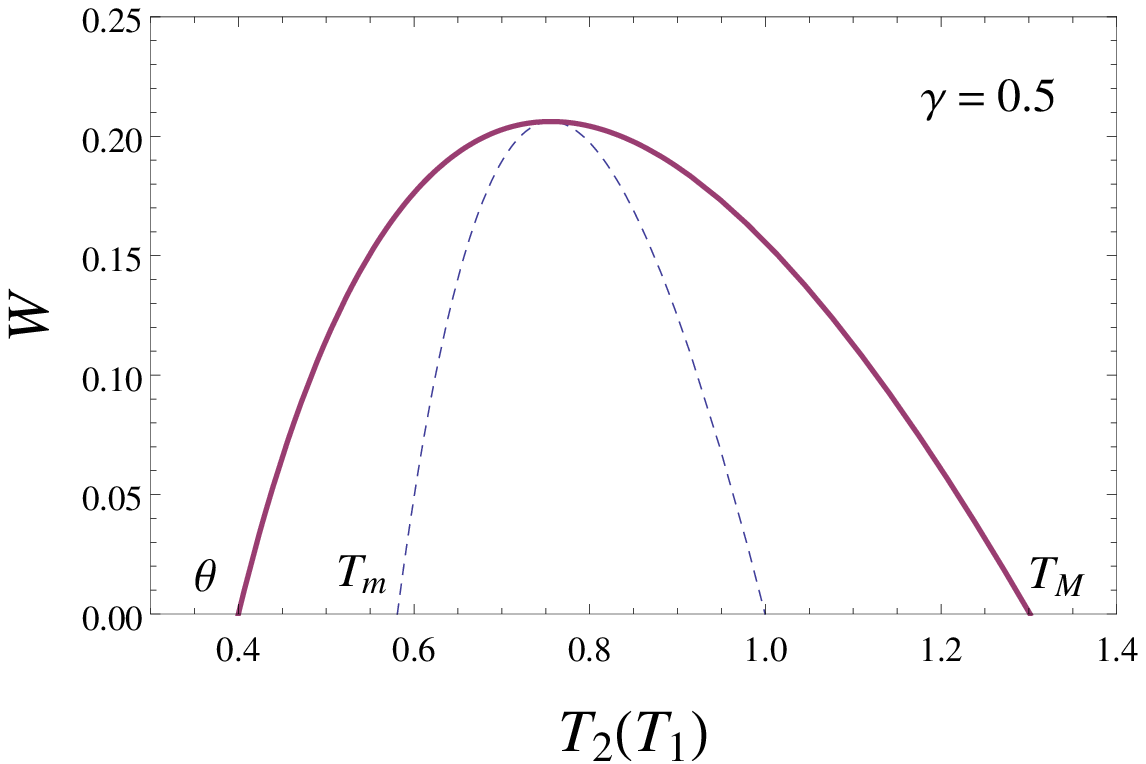} & \includegraphics[scale = .5]{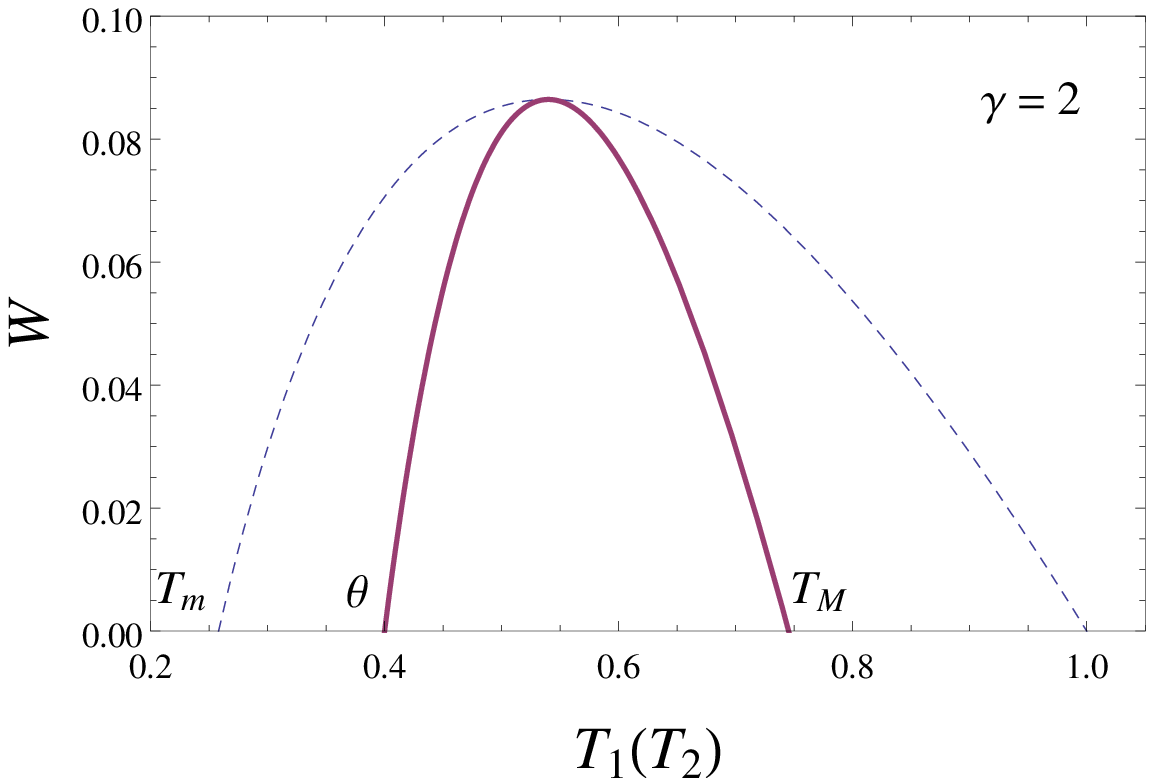}
\end{tabular} 
\caption{\it Work as a function of $T_1$ (Dashed Curve) and $T_2$ (Solid Curve) for $\omega_1 = 0.1$, $\theta = 0.4$}.
\label{figrange}
\end{figure}

With identical systems, we had an additional assumption of the same form of normalised prior distribution, $P$, for $T_1$ and 
$T_2$ and thus prior distribution is invariant under the change of parameter. Here, with 
non-identical systems, still we assume the same functional form $f$ of the prior distributions for $T_1$ and $T_2$. However, 
since now the two systems are not identical, this information has to be incorporated while assigning the prior. 
So, this information is incorporated in the allowed values of the range of $T_1$ and $T_2$. Now, the allowed range for
$T_1$ and $T_2$ is not [$\theta, 1$]. It will be different for both the parameters for different values of $\gamma$ ($\neq 1$).
Thus, say, $T_1$ ranges in [$T_m,1$] and $T_2$ ranges in [$\theta,T_M$] respectively satisfying the constraint $W\geq 0$ 
\cite{johaljnet}. 
This can be shown graphically in Figure \ref{figrange} for the two cases with $\gamma < 1$ and $\gamma >1$.
For $\gamma<1$ (larger source in comparison to sink), the range of allowed values of $T_1$ is narrower than the range for $T_2$. 
In the limiting case of $\gamma\rightarrow 0$ (infinite source and finite sink), [$T_m,1$] shrinks to a point $T_1 =1$ which is 
expected for an infinite source as now the temperature of the source stays at $T_+ = 1$. Similarly, for $\gamma>1$ 
(larger sink in comparison to source), the range for $T_2$ shrinks in comparison to the range for $T_1$, and [$\theta,T_M$] shrinks 
to a point $T_2 = \theta$ for $\gamma\rightarrow \infty$ (infinite sink and finite source). This information on the range of 
uncertain parameters will be incorporated to determine the normalisation constants for prior distributions and thus we can write 
$P_1$ and $P_2$ as \cite{johaljnet}:
\bea
P_1 (T_1)& = & \frac{f(T_1)}{\int_{T_m}^{1} f(T_1) dT_1},\label{priorT1}\\
P_2 (T_2)& = & \frac{f(T_2)}{\int_{\theta}^{T_M} f(T_2) dT_2}\label{priorT2},
\eea
where the form of $f$ is determined from the constraint condition $dS =0$, which can be written as:
\be
dS_1 + dS_2 = 0,
\ee
which can be further written, using additivity of energy, as:
\bea
\left(\frac{\partial S_1}{\partial U_1}\right)_V{\left(\frac{\partial U_1}{\partial T_1}\right)}_V dT_1 +
\left(\frac{\partial S_2}{\partial U_2}\right)_V{\left(\frac{\partial U_2}{\partial T_2}\right)}_V dT_2 & = & 0.
\eea
Using $(\frac{\partial S}{\partial U})_V = \frac{1}{T}$ and ${(\frac{\partial U}{\partial T})}_V = C (T)$ in above equation, 
we get:
\bea
\frac{dT_1}{dT_2} & = & -\frac{C_2/T_2}{C_1/T_1},
\label{inf}
\eea
Combining Eq.(\ref{inf}) and Eq.(\ref{priorrel}), suggests the following form of prior \cite{preetyarxv}:
\be
P(T_i)=\frac{C_i(T_i)/T_i}{N},
\label{pr}
\ee
where $i=1,2$ and $N=\int_{\theta}^{1} C_i(T_i)/T_i \;dT_i$ is the normalisation constant. With our model of 
the reservoirs ($S_i\propto {U_i}^{\omega_1}$), the functional form of the prior distribution can be written 
as:
\be
P(T_i) \propto {T_i}^{\omega-1}.
\label{fx2}
\ee
\section{Estimation of temperature}
The expected value of a temperature is:
\be
\overline{T_i} = \int_{T_{i,min}}^{T_{i,max}} T_i P(T_i) dT_i,
\ee
where $i=1,2$. Taking into account the respective ranges of allowed values of $T_1$ and $T_2$, identified above, 
we obtain:
\be
\overline{T}_1 = \left (\frac{\omega}{1+\omega}\right)\left(\frac{1-{T_m}^{1+\omega}}{1-{T_m}^{\omega}}\right),
\label{avT1}
\ee
and
\be
\overline{T}_2 = \left (\frac{\omega}{1+\omega}\right)\left(\frac{{T_M}^{1+\omega}-{\theta}^{1+\omega}}
{{T_M}^{\omega}-{\theta}^{\omega}}\right).
\label{avT2}
\ee
To determine $T_m$ or $T_M$, we solve Eq. (\ref{work}) by setting $W(T_1) = 0$ or $W(T_2) = 0$ respectively.
In general, Eq. (\ref{work}) has to be solved numerically for arbitrary values of $\omega_1$.
For ideal Fermi gas ($\omega_1 = 1/2$), it can be solved analytically \cite{katyayan} and thus we get:
\bea
T_1 &\in&\left[\frac{1-\gamma^2+2 \gamma^2\theta}{1+\gamma^2},1\right],\\
T_2 &\in&\left[\theta,\frac{2-\theta+\gamma^2\theta}{1+\gamma^2}\right].
\eea
Due to Eq. (\ref{T1rel}), we can write one-to-one relation between $T_m$ and $T_M$ as:
\be
1- {T_m}^{\omega} = \sigma ({T_M}^{\omega}-{\theta}^{\omega}).
\label{avrel1}
\ee
Using above equation in $W(T_M) = 0$, we obtain:
\be
1-{T_m}^{1+\omega} = \sigma ({T_M}^{1+\omega}-{\theta}^{1+\omega}).
\label{avrel2}
\ee
From Eqs. (\ref{avT1}), (\ref{avT2}), (\ref{avrel1}) and (\ref{avrel2}), we can write:
\be
\overline{T}_1 = \overline{T}_2.
\ee
However, firstly we will solve the Eq. (\ref{work}) for the limiting cases when one of the systems become very large in 
comparison to the other system.
\subsection{Infinite source and finite sink}
This case corresponds to the limit $\gamma\rightarrow 0$. Here the only uncertain parameter
is $T_2$ as temperature of source stays at $T_+ = 1$ while the temperature of sink approaches $T_+$ at optimal work 
extraction. To discuss this limit, we set Eq. (\ref{workT2}) as $W(T_2) = 0$ and obtain:
\be
{T_2}^{1+\omega}-{\theta}^{1+\omega} = \frac{1}{\sigma}\left[1-(1 + \sigma (\theta^{\omega} - 
{T_2}^{\omega}))^{\frac{1+\omega}{\omega}}\right].
\ee
Taking the limit $\gamma\rightarrow 0$, the above equation gets simplified to :
\be
\omega \left({T_2}^{1+\omega}-{\theta}^{1+\omega}\right) = 
(1+\omega)\left({T_2}^{\omega}-{\theta}^{\omega}\right),
\ee
whose trivial solution is $T_2 = \theta$. The other solution is $T_M$ so we write:
\be
\omega \left({T_M}^{1+\omega}-{\theta}^{1+\omega}\right) = 
(1+\omega)\left({T_M}^{\omega}-{\theta}^{\omega}\right).
\label{limT2}
\ee
Consistency between Eqs. (\ref{avT2}) and (\ref{limT2}) demands that we must have:
\be
\overline{T}_2 = 1.
\label{avlmT2}
\ee
Thus expected sink temperature exactly matches with temperature of heat source for optimal process. The efficiency is estimated 
by replacing $T_2$ in Eq. (\ref{effg0}) by Eq.(\ref{avlmT2}) and estimate for efficiency is same as Eq.(\ref{effopg0}). Hence, 
inference approach reproduces the optimal behaviour exactly in the limit $\gamma\rightarrow 0$.
\subsection{Finite source and infinite sink}
Consider the case of infinite sink in comparison to source ($\gamma\rightarrow \infty$).
Here, the sink stays at temperature $T_- (=\theta)$ and the temperature of source approaches $T_-$ for optimal work extraction. 
Hence $T_1$ is the only uncertain parameter for this limiting case. The range for $T_1$ is determined by using (\ref{T2rel}) 
in (\ref{work}) and then setting $W(T_1) = 0$,
we get:
\be
1 - {T_1}^{1 + \omega} = \sigma \left[{(\theta^{\omega} + \sigma^{-1}(1-{T_1}^{\omega}))^{\frac{1+\omega}{\omega}}
-{\theta}^{1+\omega}}\right].
\ee
In the limit $\gamma \rightarrow \infty$, the above equation gets simplifies to:
\be
\omega \left(1 - {T_1}^{1 + \omega}\right) = \theta (1 + \omega) \left(1 - {T_1}^{\omega}\right).
\ee
The trivial root of above equation is $T_1 = 1$ and other root ($T_m$) satisfies:
\be
\omega \left(1 -{T_m}^{1 + \omega}\right) = \theta (1 + \omega) \left(1 - {T_m}^{\omega}\right).
\label{limT1}
\ee
From Eqs. (\ref{avT1}) and (\ref{limT1}), we obtain:
\be
\overline{T}_1 = \theta.
\label{avlmT1}
\ee
It is clear from the Eq. (\ref{avlmT1}) that the average temperature of the source exactly matches with the temperature
of the infinite sink which happens in case of maximum work extraction. Further, efficiency at optimal work ($\eta_{\infty}^{*}$)
is also inferred exactly due to Eqs. (\ref{effgl}) and (\ref{avlmT1}).
Thus, we are able to infer exactly the optimal behaviour of the system with infinite sink 
and finite source also.
\subsection{Near-equilibrium estimation}
In this 
section, we approximate the values of $T_m$ and $T_M$ when $\theta$ is close to unity. For this, consider the case 
$0< \gamma <1$. Let us 
examine the case of observer 2. Since close to equilibrium, $T_M$ is also close to unity so we can introduce a small 
parameter $\epsilon > 0$ such that $T_M = \theta~(1 + \epsilon)$. Rewriting the Eq. (\ref{workT2}) as $W(T_M) = 0$:
\be
1 + \sigma \theta^{1+\omega} [1-(1+\epsilon)^{1+\omega}] = \left(1 + \sigma 
\theta^{\omega}[1-(1+\epsilon)^{\omega}]\right)^{\frac{1+\omega}{\omega}}.
\label{solTM}                                                     
\ee
Making series expansion in $\epsilon$ and keeping terms only upto second order, we get a quadratic equation in $\epsilon$ as:
\bea
 &&(1-\omega)\left[\sigma^{2} \theta^{2 \omega} + 3 \sigma \theta^{\omega} 
-\omega(1-\theta)+2 \right]{\epsilon}^2 \nonumber\\
&&- 3 \left[\sigma-\omega(1-\theta)+1 \right] \epsilon + 6 (1-\theta)=0.
\label{quad}
\eea
For instance, if we take limit $\omega \rightarrow 0$ in above equation, we reproduce the case for perfect gas as:
\be
(\gamma+1)(\gamma+2) \epsilon^2 - 3 (\gamma+1)\epsilon + 6 (1-\theta)= 0,
\label{quadig}
\ee
whose acceptable solution \cite{johaljnet} is approximated upto second order in $\eta_c$ as:
\be
\epsilon = \frac{2}{1 + \gamma}{\eta_c} + \frac{4(2 + \gamma)}{3 (1 + \gamma)^2}{\eta}_{c}^2.
\label{solig}
\ee
Similarly, Eq. (\ref{quad}) can be solved for $\epsilon$ and the solution can be approximated as:
\be
\epsilon = \frac{2}{1+\sigma}{\eta_c} + \frac{2 [4 - \omega + \sigma (\omega + 2)]}
{3 (1+\sigma)^2}{{\eta}_{c}^2}.
\label{sol} 
\ee
Then $T_M$ is determined, which in turn determines $T_m$ due to Eq. (\ref{T1rel}).

Suppose $\tilde{T}_1$ ($\tilde{T}_2$) are the estimates for $T_1$ ($T_2$) by the observer 2 (1) by making use of
(\ref{T1rel}) and (\ref{T2rel}). Let us examine the near-equilibrium expansion of the estimates of 
temperature as well as $T_c$ as:
\bea
\overline{T}_2 & = &1 - \frac{\sigma}{1+\sigma}\eta_c 
-\frac{\sigma (1-\omega)}{3 (1+\sigma)^2}
 {\eta}_{c}^2 + O[{\eta}_{c}^3],\\
 \nonumber\\
\tilde{T}_2 & = &1 - \frac{\sigma}{1+\sigma}\eta_c 
 -\frac{(1+ 3 \sigma) (1-\omega)}{6 (1+\sigma)^2 }
 {\eta}_{c}^2 + O[{\eta}_{c}^3],\\
 \nonumber\\
 \tilde{T}_1 & = &1 - \frac{\sigma}{1+\sigma}\eta_c 
 -\frac{\sigma (3+ \sigma) (1-\omega)}{6 (1+\sigma)^2 }
 {\eta}_{c}^2 + O[{\eta}_{c}^3],\\
 \nonumber\\
 T_c & = &  1 - \frac{\sigma}{1+\sigma}\eta_c 
 -\frac{\sigma (1-\omega)}{2 (1 +\sigma)^2}
 {\eta}_{c}^2 \nonumber \\
 &-&\frac{\sigma (2+\sigma+\omega(\sigma-1))(1-\omega)}
 {6 (1+\sigma)^3}{\eta}_{c}^3 + O[{\eta}_{c}^4].
\eea
We have skipped the lengthy expressions of third order terms in series expansion of temperature estimates.
Let us define the estimated value of the temperature of one reservoir as the weighted mean of the 
estimates by two observers. It will be given as
\be
T_{i,m} = A \overline{T}_i + B \tilde{T}_i,
\ee
where $i=1,2$ and $A$, $B$ are the weights satisfying the condition $A + B = 1$. In the above case, we have seen that both
estimates, $\overline{T}_2$ (by observer 2) and $\tilde{T}_2$ (by observer 1), match with $T_c$ only upto 
first order. If we choose weights $A$ and $B$ so as to obtain matching beyond first order, then the weights are
calculated as:
\bea
A &=& \frac{1}{1+\sigma},\\
B &=& \frac{\sigma}{1+\sigma}.
\eea
This weighted mean estimated temperature ($T_{2,m}$) of the sink shows remarkable agreement with $T_c$ upto third order 
in $\eta_c$ close to equilibrium. Further, for $\gamma\rightarrow 0$, $T_{2,m}$ becomes exactly equal to $\overline{T}_2$
showing that estimation is done only by observer 2 as the other reservoir corresponding to observer 1 (source) becomes 
infinite in comparison and hence, its temperature stays constant at $T_+$ and no uncertainty exists in its value as the 
process proceeds. In other limit of $\gamma \rightarrow \infty$,  $\overline{T}_1$ becomes exactly equal to $T_c$.

\section{Estimation of efficiency}
Efficiency is estimated by replacing $\overline{T}_2$ in Eqs. (\ref{T1rel}) and (\ref{workT2}) to obtain efficiency estimate by 
observer 2 ($\tilde{\eta_2}$). Similarly, denote the efficiency estimate by observer 1 as ($\tilde{\eta_1}$).
Expanding the estimates of efficiency close to equilibrium and make a comparison with the efficiency at optimal
work as follows:
\bea
\tilde{\eta}_1 &=& \frac{\eta_c}{2} + \frac{2(1+\sigma)+\omega(\sigma-2)}
{12 (1+\sigma)}{\eta}_{c}^2 + O[{\eta}_{c}^3],\\
\tilde{\eta}_2 &=&\frac{\eta_c}{2} + \frac{(1+\sigma)+\omega(2\sigma-1)}
{12 (1+\sigma)}{\eta}_{c}^2 + O[{\eta}_{c}^3],\\
\eta_{\gamma}^{*} & = & \frac{\eta_c}{2}+\frac{(1 + 2 \sigma)+ \omega (\sigma-1)}
{12 (1+\sigma)}{\eta}_{c}^2 + O[{\eta}_{c}^3].
\eea
It is clear from the above expressions that estimates of efficiency either by observer 1 ($\tilde{\eta}_1$) or by observer 2 
($\tilde{\eta}_2$) matches with efficiency at optimal work ($\eta_{\gamma}^{*}$) only upto first order in $\eta_c$. However, 
we define mean efficiency ($\tilde{\eta}_m$) and compare it with $\eta_{\gamma}^{*}$ as:
\bea
\tilde{\eta}_m = A \tilde{\eta}_1 + B \tilde{\eta}_2,
\eea
where $A$ and $B$ are the weights assumed for the estimation done by observer 1 and observer 2 respectively satisfying
$A + B = 1$, similar to the case of temperature estimation. Estimated mean efficiency ($\tilde{\eta}_m$) matches with
efficiency at optimal work upto second order.
For the limiting cases of $\gamma\rightarrow 0$ and $\gamma \rightarrow \infty$, $\tilde{\eta}_2$ and $\tilde{\eta}_1$
give the exact estimates for efficiency at optimal work respectively.


\subsection{Numerical results for arbitrary $\gamma$}
\begin{figure}[h]
\begin{tabular}{ll}
\includegraphics[scale=0.5]{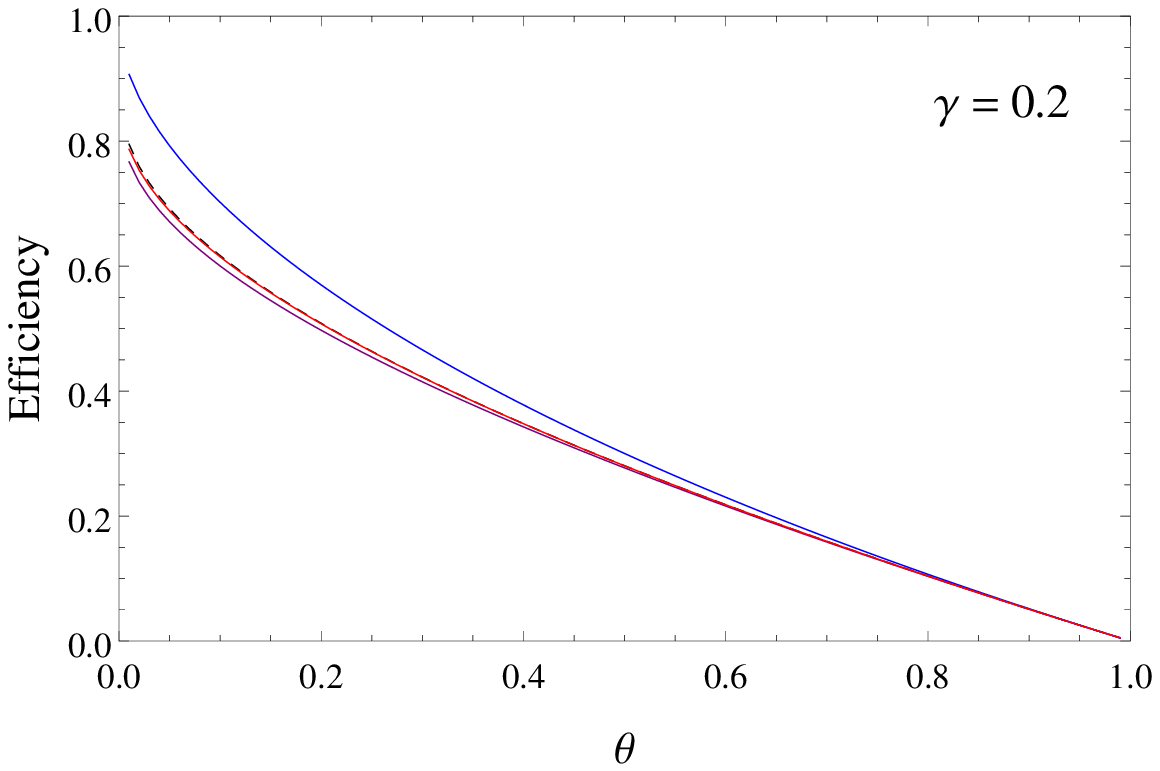}&\includegraphics[scale=0.5]{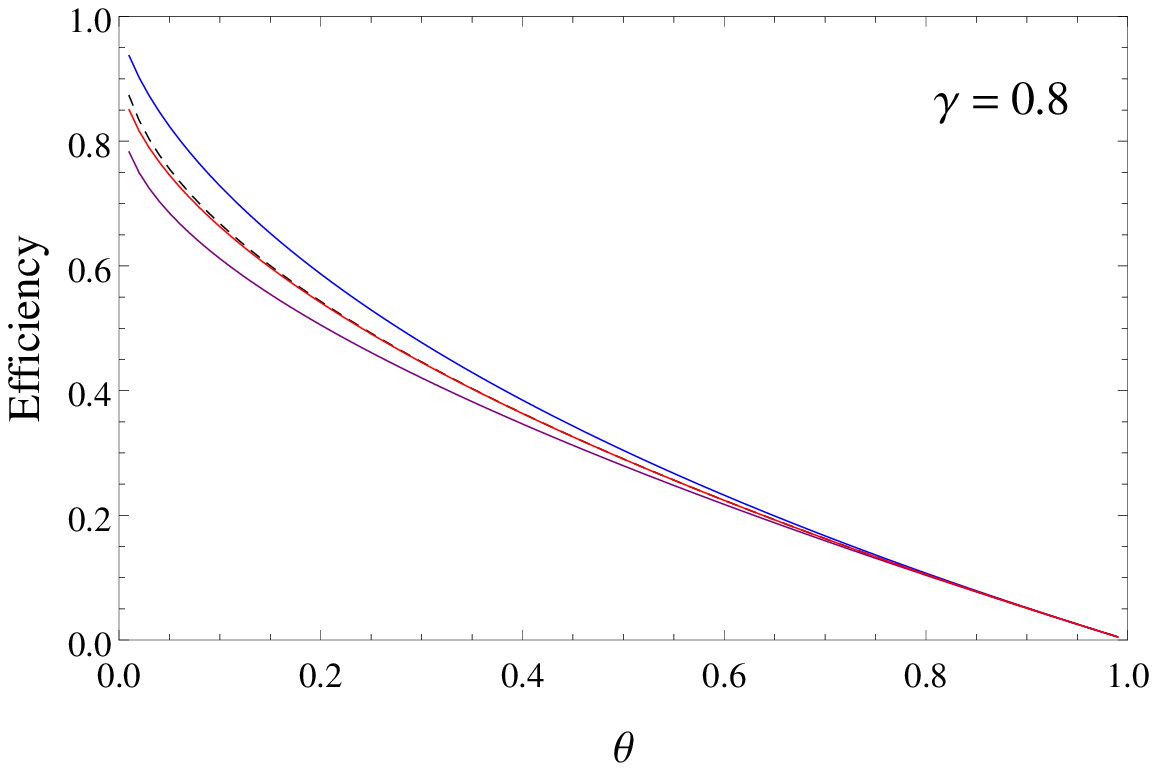}\\
\includegraphics[scale=0.5]{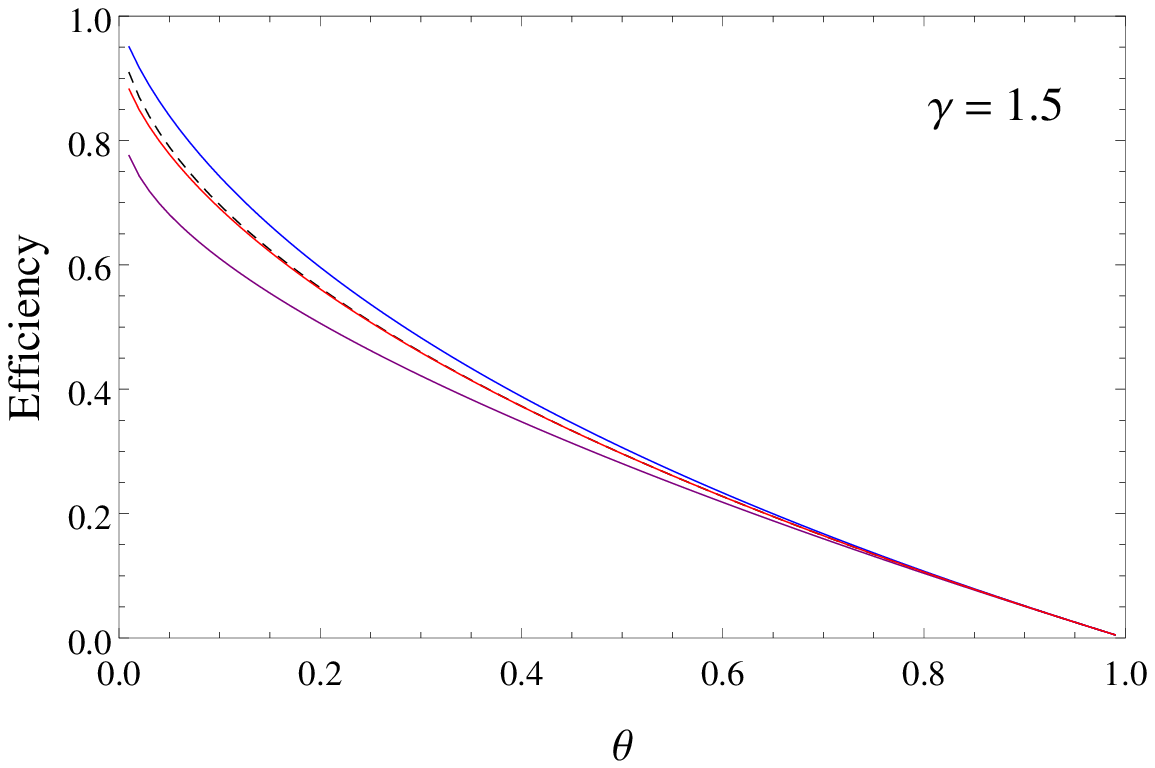} &\includegraphics[scale=0.5]{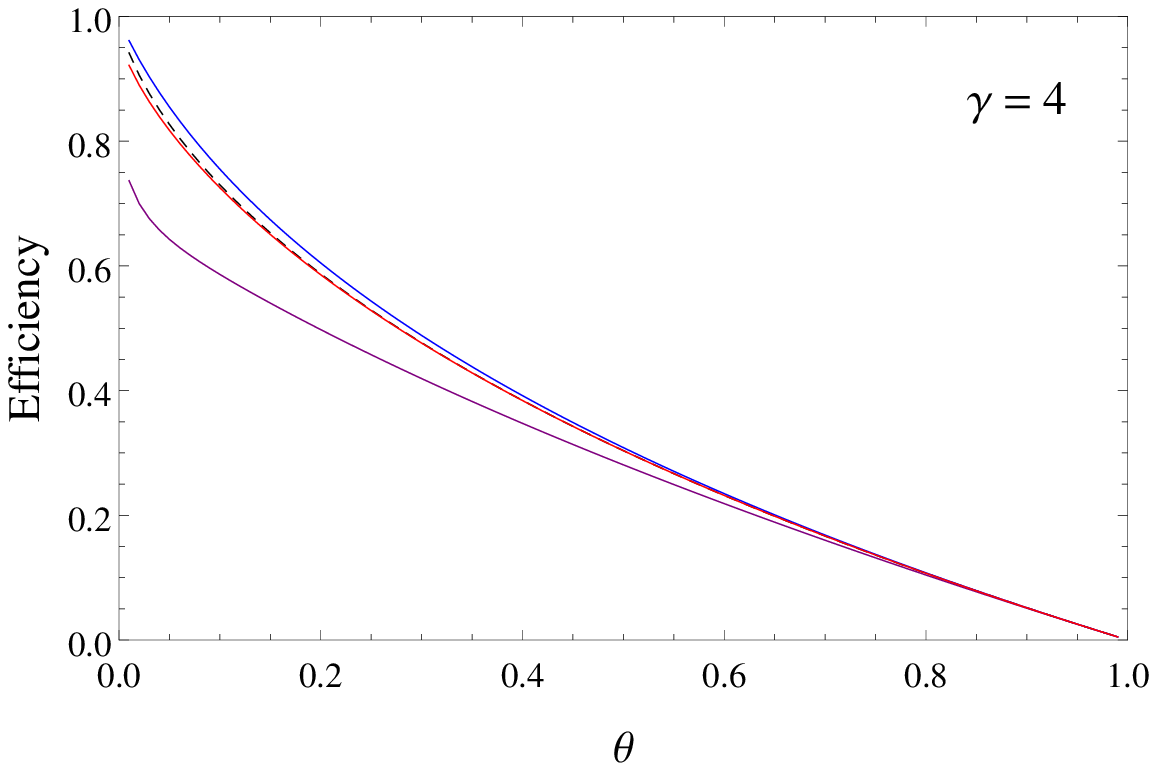}\\
\end{tabular}
\caption{\it Efficiency as a function of $\theta$ for $\omega_1 = 0.1$. The dashed curve is for efficiency at optimal work, 
uppermost curve (blue online) is for $\tilde{\eta}_1$, lowermost curve (purple online) is for $\tilde{\eta}_2$, and middle 
curve (red online) is for mean efficiency ($\tilde{\eta}_m$).}
\label{effplot}
\end{figure}

{\label{nresults}}
Eq. (\ref{workT2}) has to be solved to determine the roots, one trivial root is $\theta$ while the other root,
$T_M$, can be determined numerically for given values of $\omega_1$, $\gamma$, and $\theta$. Then, we obtain numerical
estimates of efficiency by observer 1 ($\tilde{\eta}_1$) and observer 2 ($\tilde{\eta}_2$) for arbitrary values
of $\gamma$. 
Figure \ref{effplot} shows 
the comparative plots of efficiency for different $\gamma$'s.

From the numerical plots, it has become clear that in the $0<\gamma<1$ regime, estimates made by 
observer 2 give better results as compared to the observer 1 and vice-versa in the case of $\gamma > 1$.

\section{Conclusion}{\label{conc4}}
Thus, we have extended our previous approach of inference studied in \cite{preetyiop} where two identical finite systems were 
taken as source and sink. 
Taking non-identical systems, we can distinguish or label the two systems acting as source and sink.
The range for $T_1$ and $T_2$ is also different because of dissimilar systems. While generalising this approach, we have observed 
that estimates match exactly with their optimal values when one of the reservoir becomes very large as compared to the other. 
For arbitrary values of $\gamma$ also, 
numerical calculations have been performed. These calculations show that information incorporated in the prior distribution 
reproduce the optimal behaviour of the system, however, now the efficiency estimates made by two observers 
are not symmetrically distributed about the efficiency at optimal work unlike in the case with similar reservoirs 
($\gamma = 1$). Instead, estimates made by one observer lie closer to the optimal value as compared to the other depending 
upon the value of $\gamma$. Near-equilibrium, it has been observed that universality, $\eta_{c}^2/8$, in efficiency does not
hold and becomes system dependent. However, in this case, efficiency at optimal work  can be reproduced upto second order
by defining mean efficiency with non-identical weights for the efficiency estimates by the two observers. Thus, with non-identical 
systems also, we quantify the prior information and use it to estimate the optimal performance in constrained thermodynamic process
\section{Acknowledgements}
RSJ acknowledges financial support from the Department of Science and Technology, India under the research project
No. SR/S2/CMP-0047/2010(G). PA is thankful to University Grants Commission, India and IISER Mohali for Research fellowship.

\end{document}